\documentclass[10pt, conference, compsocconf]{IEEEtran}
\usepackage{booktabs,amsmath}

\usepackage{pifont}
\usepackage{amssymb}
\usepackage[pdftex]{graphicx}
\usepackage{amsmath}
\usepackage{caption}
\usepackage{tabularx}
\usepackage{bbm}
\usepackage{ amssymb }

\ifCLASSINFOpdf
\else
\fi
\begin{document}
\title{Connection Discovery using Shared Images by Gaussian Relational Topic Model}


% \author{\IEEEauthorblockN{Xiaopeng Li}
% \IEEEauthorblockA{Dept. Electronic and Computer Engineering\\
% The Hong Kong University of Science and Technology\\
% Hong Kong\\
% Email: xlibo@connect.ust.hk}
% \and
% \IEEEauthorblockN{Ming Cheung}
% \IEEEauthorblockA{Dept. Electronic and Computer Engineering\\
% The Hong Kong University of Science and Technology\\
% Hong Kong\\
% Email: cpming@ust.hk}
% \and
% \IEEEauthorblockN{James She}
% \IEEEauthorblockA{Dept. Electronic and Computer Engineering\\
% The Hong Kong University of Science and Technology\\
% Hong Kong\\
% Email: eejames@ust.hk}
% }
\author{
    \IEEEauthorblockN{Xiaopeng Li\IEEEauthorrefmark{1}, Ming Cheung\IEEEauthorrefmark{2}, James She\IEEEauthorrefmark{3}}\\
    \IEEEauthorblockA{\IEEEauthorrefmark{1}\IEEEauthorrefmark{2}\IEEEauthorrefmark{3}HKUST-NIE Social Media Lab, Hong Kong University of Science \& Technology, Hong Kong\\
      \IEEEauthorrefmark{1}xlibo@connect.ust.hk,
      \IEEEauthorrefmark{2}cpming@ust.hk,
      \IEEEauthorrefmark{3}eejames@ust.hk}
}

%Telephone: (800) 555--1212, Fax: (888) 555--1212}


\maketitle

\begin{abstract}
Social graphs, representing online friendships among users, are one of the fundamental types of data for many applications, such as recommendation, virality prediction and marketing in social media.
However, this data may be unavailable due to the privacy concerns of users, or kept private by social network operators, which makes such applications difficult.
Inferring users' interests and discovering users' connections through their shared multimedia content has attracted more and more attention in recent years.
This paper proposes a Gaussian relational topic model for connection discovery using user shared images in social media.
The proposed model not only models users' interests as latent variables through their shared images, but also considers the connections between users as a result of their shared images. It explicitly relates user shared images to user connections in a hierarchical, systematic and supervisory way and provides an end-to-end solution for the problem.
This paper also derives efficient variational inference and learning algorithms for the posterior of the latent variables and model parameters. 
It is demonstrated through experiments with over 200k images from Flickr that the proposed method significantly outperforms the methods in previous works. 
\end{abstract}


% \fi
\IEEEpeerreviewmaketitle

\section{Introduction}
Social graphs (SGs), representing online friendships among users, are fundamental data for many applications, such as recommendation, virality prediction and marketing in social media.
However, this data may be unavailable due to the privacy concerns of users, or kept private by social network operators, and these applications become challenging with an incomplete set of data.
Providing a potential solution to this problem, user connections are also reflected in the abundant social content, especially images, shared on social networks.
Inferring users' interests and discovering users' connections through their shared multimedia content has attracted more and more attention in recent years.
%Two users with a connection will share characteristics, be follower/followee, or be member of the same community. 
%Through discovering user connections from their shared images, it is possible to identify user gender, or recommend followers/followees to users \cite{cheung2015connection}.
%One of the possible ways to do this is using the user annotation tags on their shared images.
%=============================
\begin{figure}
  \centering
  %\vspace{-10pt}
  \includegraphics[width=3.5in]{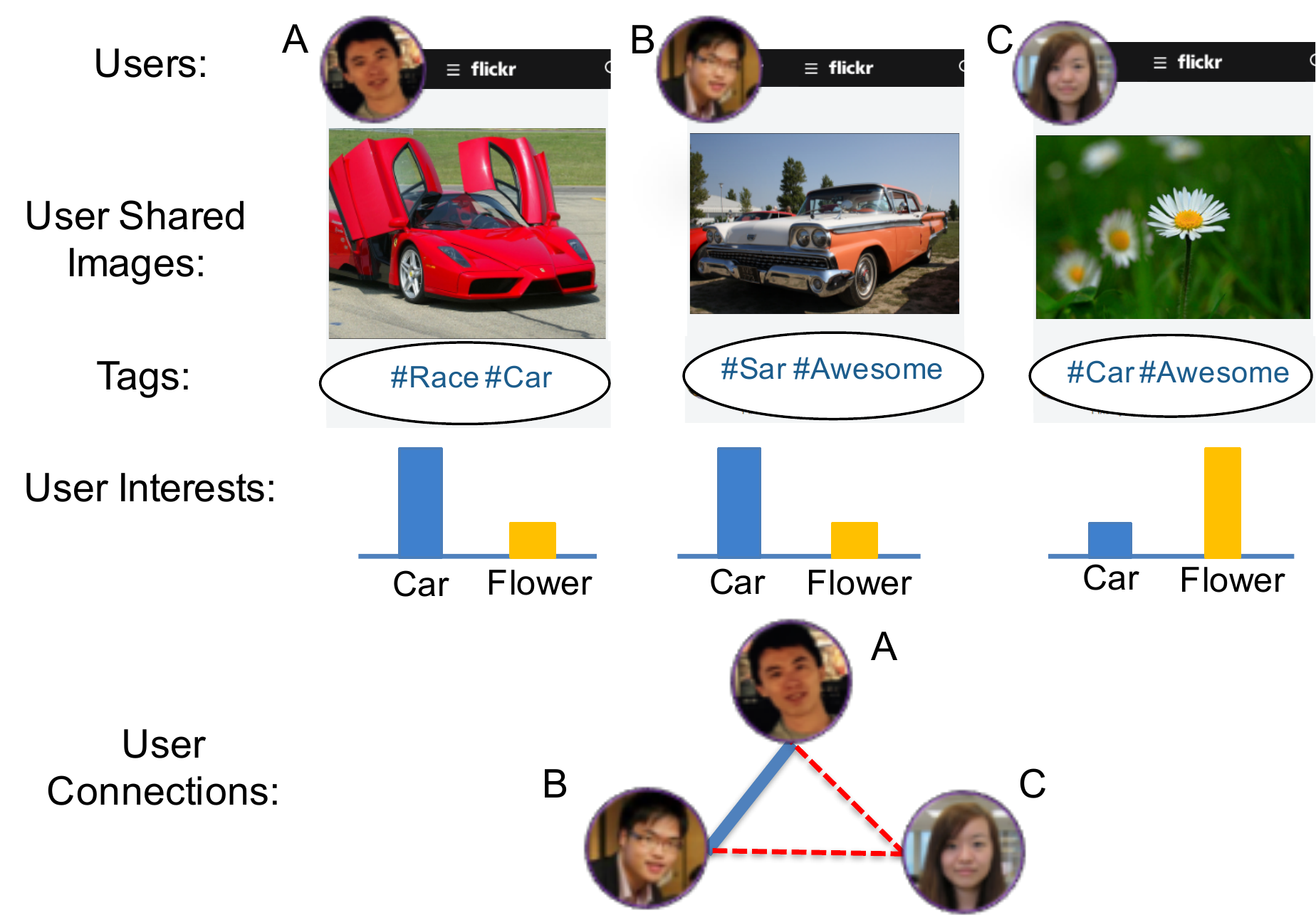}
  \caption{Examples of user shared images, image tags, user interest reflected and user connections.}
  \label{fig:interestsImageFeature}
\end{figure}
%=================================
% the major interest of social media analysis is rapidly shifting from text-based content to multimedia.
A common but unreliable approach is using user annotated tags (or user tagging) associated with each shared image to discover user connections when the SG is not accessible. However, user annotated tags may be unavailable or images may be incorrectly labeled, as shown in Fig. \ref{fig:interestsImageFeature}. 
Instead of indirectly taking the image content information through tags, a more direct method is to consider the image's visual content. 
Users with connections of follower/followee relationships are found to have relatively higher visual content similarities among their shared images. 
An simplistic example of user generated images on Flickr is shown in Fig. \ref{fig:interestsImageFeature}: Both users $A$ and $B$ share images of cars and user $C$ shares an image of a flower.
The follower/followee relationship between users $A$ and $B$ can possibly be detected from the higher similarity of visual features in their shared images. 
When more shared images from each of users $A$, $B$ and $C$ are accessible for evaluation, the actual follower/followee relationships should become reliably and accurately detectable, though such a task is becoming challenging with the number of shared images and user connections in social networks growing larger every day. 

The effectiveness of connection discovery using user shared images is mainly determined by two elements: effective extraction of information from images and an effective method to connect the image content to user connections. On the one hand, with the recent development of convolutional neural network (CNN), the analysis and understanding of image content has become much more effective \cite{lecun2015deep} through hierarchical representation, closing the semantic gap between pixels and content. It is therefore possible to extract rich image content information through CNN.
%Feature selection problem for the connection discovery using user shared images as posed by \cite{tagging-cpscom2015} is no longer a big issue. 
On the other hand, how to effectively discover user connections using the image content remains a challenge.
Summarizing the methods or frameworks proposed in previous works such as \cite{cheung2015connection}, \cite{yang2015beyond}, \cite{zhuang2011modeling} and \cite{geng2014one}, the process is generally divided into two stages: first, construction of user profile by counting the occurrences of image labels or summing the image feature vectors, and second, prediction of connections between users based on the constructed user profile. The limitation of these methods is two-fold. First, aggregating all items in a simple way for each user might not be solid enough to capture users' interests. An analogy can be found in text analytics, where the bag-of-words model is compared with topic models, especially latent Dirichlet allocation (LDA)\cite{blei2003latent}. Second, user profiling through images is completely separated from connection discovery. Separating the two potentially results in untight relatedness of image content and social links, meaning that the observation of partial social links lends no help toward connection discovery using image content.

In this paper, a Gaussian relational topic model (GRTM) is proposed for connection discovery using user shared images in order to overcome the abovementioned limitations. GRTM is an end-to-end hierarchical model specifically designed to not only model users' interests through image content but also to supervise the modeling in such a way that the content of shared images is statistically connected to the links between users, inspired by hierarchical relational model in text document domain \cite{chang2009relational}. GRTM effectively extracts rich image content information through CNN and models each semantic topic as a Gaussian topic. GRTM also models each user's interests as a latent factor and assumes that the action of the user sharing an image is probabilistically motivated by his or her interests. Furthermore, the links between users are modeled based on the images each user shares. Combining these in a coherent probabilistic generative process, the proposed GRTM provides a systematic way to close the gap between the actions of users sharing images and users connecting to each other.
The main contributions of this paper are the following:
\begin{itemize}
\item proposes an end-to-end Gaussian relational topic model for connection discovery using user shared images, closely relating user shared image content to user connections.
\item derives efficient variational inference for the proposed model to approximate the posterior of latent variables and learn model parameters.
\item evaluates the performance of the proposed model with real data and proves the significantly better performance of the proposed method.
\end{itemize}

The rest of the paper is organized as follows. Section II discusses previous works.
Section III introduces the GRTM for connection discovery using shared images. 
Section IV describes in detail the inference of latent variables and model parameters, as well as the prediction method. Section V presents experimental results and discussion, and Section VII concludes the paper.

\section{Related Works}
%start talking about SNA approaches for friendship recommendation
% As discussed above, one of the possible ways to understand images for connection discovery is through user annotated tags, a common feature in social networks, that are text-based notations of social media content.
% However, such tags can be unreliable, and the performance of the discovered connections can be affected \cite{cheung2015connection}.
%============talk about image-based methods
% Another possible way of discovering connections is a content-based approach, in which the visual elements of an image are considered in order to generate a label for the image \cite{moxley2009not}\cite{zhang2013social}.
One content-based approach to discover connections through images is to generate a label for an image based on its visual elements \cite{moxley2009not}\cite{zhang2013social}.
However, determining the relationship between the visual elements and the label is not a trivial task because the same object can be visually different among images, and denoting each image by a single label loses a lot of information.
\cite{cheung2015connection} and \cite{tagging-cpscom2015} propose to first generate labels for user shared images by clustering and describe users' characteristics by counting the occurrences of image clusters appearing in their collections. The prediction is then made through analyzing the similarities among the users' histograms. Both methods assume each image contains one and only one topic, neglecting potentially rich information. On the other hand, \cite{yang2015beyond} and \cite{zhuang2011modeling} represent users' interests by summing the feature vectors of their images, achieving a similar effect to the other methods. However, all these methods do not actually connect shared images to the links between users, or only connect shared images to links, and not the other way around. 
% The issue is that with evidence of observed links, there is no way to guide the framework to make better predictions using images. Therefore, the methods might not be effective in using shared images for prediction.

% Previous works, such as \cite{yuan2013latent}, have developed relational deep learning model (deep belief nets) for the link analysis problem in social media netoworks. However, the methods generally do not apply to the problem of connection discovery using shared images since the latent features are learned from the raw feature of one entity, instead of a collection of image features. 
% %Preprocessing the image features by Bag of Features potentially loses rich information contained in the images.
% \cite{geng2014one} shares a similar philosophy to this paper. It considers both image content and social information to construct user profiles for further applications.
% % where the paper first constructs profile ontology for images using Multi-task CNN and obtain user profile by averaging users' image concepts then uses social information to refine the user profile. 
% However, the construction of the ontology used in the paper explicitly requires user curated training images, with the images well-organized and the concept content known.

Relational topic model was first proposed in \cite{chang2009relational} for document networks in the text analytics domain. It is an extension of the latent Dirichlet model (LDA) \cite{blei2003latent} and supervised LDA \cite{mcauliffe2008supervised}. This family of approaches model each document as topic proportions, and each word in the document is drawn from a set of topics, which are distributions over a fixed vocabulary. Furthermore, the relational topic model models the links between documents as a binary random variable that is conditioned on their contents.
More recent papers, \cite{xie2015mining} and \cite{sang2012right}, share a similar methodology with this paper, also proposing a hierarchical topic model to process user images and infer users' interest as latent factors from Bayesian inference. However, \cite{xie2015mining} starts with low-level pixels and describes image regions with visual words, while \cite{sang2012right} models the visual descriptor as drawn from a vocabulary, just like in the text domain. Also, \cite{xie2015mining} considers no social information and hence, as with other works, does not relate image content to user relationships for connection discovery, and \cite{sang2012right} considers the social influence of existing links to the users rather than predicting new ones.
Our work extracts rich information of image content through CNN and models users' interests and users' connections simutaneously. As an end-to-end model, ours closely relates users' connections to user shared images, with the goal being connection discovery.

\section{The proposed model for Connection Discovery using Shared Images}
\begin{figure}
  \centering
  %\vspace{-10pt}
  %\includegraphics[width=6.5in]{image/overview.png}
  \includegraphics[width=2.5in]{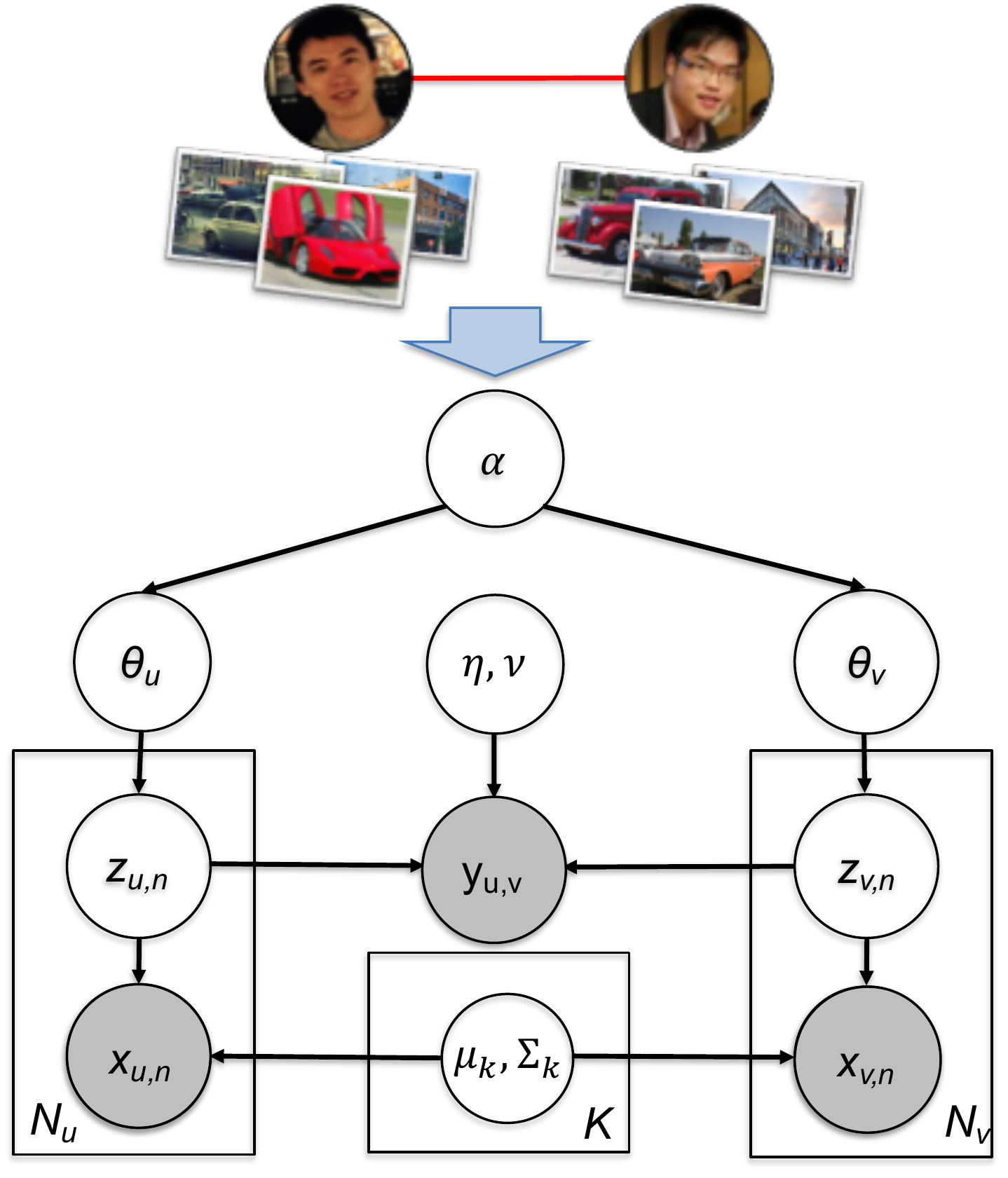}
  \caption{The proposed Gaussian Relational Topic Model (GRTM) (shaded nodes are observed.)}
  \label{fig:systemFlow}
\end{figure}
This section introduces the problem of connection discovery using shared images and the proposed Gaussian relational topic model.
Given a social network, there are $N$ users, $\{u_1,u_2,\dots,u_N\}$. Each of the users, $u$, shares a collection of images, $I_u = \{x_{u,1},x_{u,2},\dots,x_{u,N_u}\}$, where $N_u$ is the number of images the user $u$ shares and $x_{u,n}$ is the $n$th image shared by the user $u$. The connections between users are not available due to privacy concerns or proprietary information protection. However, the connections can still be discovered through the images shared by the users. The objective is to predict the possible connections between users based on the user shared images.

\begin{figure}
  \centering
  %\vspace{-10pt}
  \includegraphics[width=3.5in]{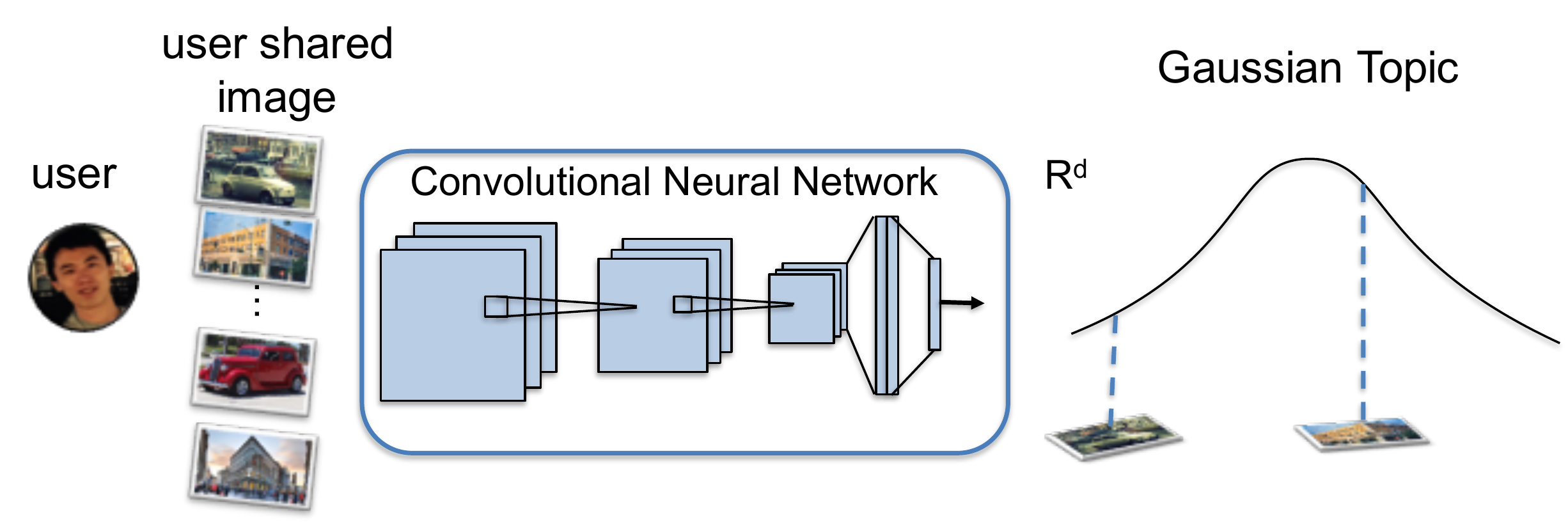}
  \caption{CNN feature extraction for images and Gaussian topic}
  \label{fig:gaussian_topic}
\end{figure}

The proposed GRTM model, is based on Gaussian mixture model and relational topic model, as shown in Fig. \ref{fig:systemFlow}. It is a generative probabilistic model in which a user is described by a topic distribution that is reflected through his or her collection of shared images. It models that the action of a user sharing an image follows a generative process: a user's preferences or the topic proportion in his or her collections is generated through a Dirichlet-distributed vector, and he shares an image by first drawing a topic assignment from his or her preferences then drawing an image from the corresponding topic distribution. 
Unlike LDA and relational topic model, where text documents are the context and words are generated from a vocabulary, this paper extracts rich information from images through a pre-trained CNN and Gaussian topic is adopted in order to consider and preserve the rich information of the images, as shown in Fig. \ref{fig:gaussian_topic}.
The links between users are then modeled as binary variables and are determined by the users' preferences. In this way, the user shared images and the links between users are statistically connected.

Formally, the generative process of the proposed GRTM is as follows:
\begin{enumerate}
  \item For each user $u$:
    \begin{enumerate}
    \item Draw topic proportions $\theta_u|\alpha \sim Dir(\alpha)$.
    \item For each image $x_{u,n}$:
      \begin{enumerate}
      \item Draw topic assignment $z_{u,n}|\theta_u \sim Mult(\theta_u)$.
      \item Draw the image $x_{u,n}|z_{u,n},\{\mu,\Sigma\}_{1:K} \sim \mathcal{N}(\mu_{z_{u,n}},\Sigma_{z_{u,n}})$.
      \end{enumerate}
    \end{enumerate}
  \item For each pair of users $u,v$:
    \begin{enumerate}
      \item Draw binary link indicator between users $y|z_u,z_v \sim \psi(\cdot|z_u,z_v)$.
    \end{enumerate}
\end{enumerate}

Fig. \ref{fig:systemFlow} illustrates the graphical model for the generative process. Assuming there are a total of $K$ topics, generating an image is done by first picking the topic it belongs to. The apparent choice of distribution for picking the topic assignment $z_{u,n}$ from $K$ possible values is multinomial distribution, parameterized by the $k$-dimensional user preference $\theta_u$. 
The conjugate prior for multinomial distribution is Dirichlet distribution. Therefore, the user preference $\theta_u$ is assumed to be drawn from Dirichlet distribution, parameterized by $\alpha$.

Under each topic, an image is assumed to be drawn from a multivariate Gaussian distribution, parameterized by mean $\mu$ and covariance $\Sigma$:
\begin{equation}
p(x_{u,n}|z_{u,n},\{\mu,\Sigma\}_{1:K}) = \prod_{k=1}^{K} \mathcal{N}(x_{u,n}|\mu_k,\Sigma_k)^{\mathbbm{1}(z_{u,n}=k)}.
\end{equation}
where $\mathbbm{1}(\cdot)$ is an indicator function. Therefore, given the parameters $\alpha$ and topic distribution $\{\mu,\Sigma\}_{1:K}$, the generative probability of a user shared image is given by
\begin{equation}
\begin{split}
p(x_{u,n},&z_{u,n},\theta_u|\alpha,\{\mu,\Sigma\}_{1:K}) \\
&= p(\theta_u|\alpha)p(z_{u,n}|\theta_u)p(x_{u,n}|z_{u,n},\{\mu,\Sigma\}_{1:K}).
\end{split}
\end{equation}

Given two users and the topic assignments of all their images, the link between the two users is determined by the link probability function $\psi$. Specifically, the exponential function is adopted due to the linear form of the log likelihood \cite{chang2009relational}:
\begin{equation}
\psi(y=1|z_u,z_v) = \exp(\eta^T(\bar{z_u} \circ \bar{z_v}) + \nu),
\end{equation}
where $\bar{z_u} = \frac{1}{N_u} \sum_n z_{u,n}$, $\eta$ and $\nu$ are parameters, and the notation $\circ$ denotes the element-wise product.

The overall joint likelihood of the observation, i.e., the user shared images and links between users, and the latent variables, i.e., the user preference and topic assignment, is determined by
\begin{equation}
\begin{split}
&p(U,y,\theta,z|\alpha,\{\mu,\Sigma\}_{1:K}) = \prod_{u}p(\theta_u|\alpha) \\
&\prod_n p(z_{u,n}|\theta_u)p(x_{u,n}|z_{u,n},\{\mu,\Sigma\}_{1:K}) \prod_{(u,v)}p(y|z_u,z_v).
\end{split}
\end{equation}

% With the likelihood available, the Bayesian method enables us to do posterior inference of the latent variables and parameters based on the evidence of observations. 
% With the user shared images and the links between users available, the latent user preference and topic assignment for each image, as well as the parameters in the link function, can be estimated. Thus, the causal loop between the actions of users sharing images and users linking to each other is closed.

\section{Inference, Estimation and Prediction}
In this section, the Bayesian variational inference for the proposed GRTM is presented. 
% The estimation of the latent variables and parameters is conducted based on the evidence of the observation through Bayesian approach. And the prediction for links between users, given the user shared images, using the estimated parameters is described.

\subsection{Inference}
The posterior distribution of the latent variables and parameters given the observations and topic parameters is inferred by:
\begin{equation}
\begin{split}
p(\theta,z,\eta,\nu &| U,y,\alpha, \{\mu,\Sigma\}_{1:K}) \\
&= \frac{p(U,y,\theta,z|\alpha,\{\mu,\Sigma\}_{1:K})}{\int \sum_{z_{u,n}} p(U,y,\theta,z|\alpha,\{\mu,\Sigma\}_{1:K}) d\theta}.
\end{split}
\end{equation}
The exact posterior, however, is intractable since the denomenator involves a complex summation and integral. Instead, like previous works, variational inference with free parameters is used to approximate the exact posterior. Specifically, the mean field approximation with the form
\begin{equation}
\begin{split}
q(\theta,z | \gamma, \phi) = \prod_{u} q_u(\theta_u | \gamma_u) \prod_n q_z(z_{u,n} | \phi_{u,n})
\end{split}
\end{equation}
is used to approximate the posterior $p$ of $\theta$ and $z$. And the goal is to minimize the Kullback-–Leibler (KL) divergence between the approximation and the exact posterior. Equivalently, we try to maximize the variational free energy, which is the evidence lower bound of the log marginal probability of the observations:
\begin{equation}
\label{eq:elbo}
\begin{split}
\mathcal{L} = &\sum_u\sum_n E_q[\log p(x_{u,n}|z_{u,n},\{\mu,\Sigma\}_{1:K})] \\
&+ \sum_u\sum_n E_q[\log p(z_{u,n}|\theta_u)] +\sum_u E_q[\log p(\theta_u|\alpha)] \\
&- \sum_u E_q[\log q_{\theta}(\theta_u | \gamma_u)] - \sum_u\sum_n E_q[\log q_z(z_{u,n}|\phi_{u,n})] \\
&+ \sum_{(u,v)} E_q[\log p(y_{u,v}|z_{u,v},\eta,\nu)].
\end{split}
\end{equation}

\subsection{Estimation}
Having the evidence lower bound of the log likelihood, the maximization of the lower bound can be achieved through coordinate ascent, where each variable is iteratively updated assuming all others are fixed, until convergence. 
% In this subsection, the update rule of each latent variable and parameter is derived. 
By taking the derivative of the evidence lower bound in Eq. \ref{eq:elbo}, subject to the sum-to-one constraint, the update of the free parameter $\phi$ is
\begin{equation}
\begin{split}
\phi_{u,n,k} \propto \exp(\log N(x_{u,n}| \mu_k, \Sigma_k) & + \Psi(\gamma_i) - \Psi(\sum_{j=1}^{K}\gamma_j) \\
&+ \sum_{v|y_{u,v}=1} \frac{\eta \circ \bar{\phi}_v}{N_u}),
\end{split}
\end{equation}
where $\bar{\phi}_v = E_q[\bar{z}_v]=\frac{1}{N_v}\sum_n\phi_{v,n}$. And the normalization term is the sum over all $K$.

With a similar method, the update of the free parameter $\gamma$ is
\begin{equation}
\gamma_{u,k} = \alpha_k + \sum_{n=1}^{N_u}\phi_{u,n,k}.
\end{equation}

Isolating the terms involving $\{\mu,\Sigma\}_{1:K}$ in the lower bound and taking the derivative, the update of the topic parameters $\{\mu,\Sigma\}_{1:K}$ is
\begin{equation}
\begin{split}
\mu_k = \frac{\sum_{u}\sum_n \phi_{u,n,k}x_{u,n}}{\sum_u\sum_n\phi_{u,n,k}},
\end{split}
\end{equation}
\begin{equation}
\begin{split}
\Sigma_k = \frac{\sum_u\sum_n\phi_{u,n,k}(x_{u,n}-\mu_k)(x_{u,n}-\mu_k)^T}{\sum_u\sum_n\phi_{u,n,k}}.
\end{split}
\end{equation}

It would be inappropriate to regard all links except observed positive links as negative training examples since there might be positive but unobserved links between users, which is expected to predict. Therefore, following \cite{chang2009relational}, we use a regularization penalty parameterized by $\rho$ for the negative observations. And the updates of the link function are conducted analytically by
\begin{equation}
\begin{split}
\nu \leftarrow \log(M-\mathbf{1}^T \bar{\Pi}) - \log(\rho(1-\frac{1}{K})+M-\mathbf{1}^T\bar{\Pi}),
\end{split}
\end{equation}
\begin{equation}
\begin{split}
\eta \leftarrow \log(\bar{\Pi}) - \log(\bar{\Pi}+\frac{\rho}{K^2}\mathbf{1} - \mathbf{1}\nu),
\end{split}
\end{equation}
where $M=\sum_{u,v} \mathbbm{1}(y_{u,v}=1)$, $\bar{\Pi} = \sum_{u,v}\bar{\pi}_{u,v}\mathbbm{1}(y_{u,v}=1)$ and $\bar{\pi}_{u,v} = \frac{1}{N_u}\sum_n\phi_{u,n} \circ \frac{1}{N_v}\sum_n\phi_{v,n}$.
\subsection{Prediction}
With all the variational parameters estimated, the link between two users given their shared collection of images can be predicted. The variational Bayesian prediction is given by:
\begin{equation}
\begin{split}
p(y_{u,v}|x_u,x_v) = E_q[p(y_{u,v}|\bar{z}_u,\bar{z}_v)].
\end{split}
\end{equation}
However, it still involves complicated summations over all possible $z$ for all the images considered. Instead, we perform plug-in approximation and substitute all $z$ with its variational expectation. Hence, the predictive probability is approximated with
\begin{equation}
\label{eq:predict}
\begin{split}
p(y_{u,v}|x_u,x_v) = \exp(\eta^T\bar{\pi}_{u,v}+\nu),
\end{split}
\end{equation}
where $\bar{\pi}_{u,v} = \frac{1}{N_u}\sum_n\phi_{u,n} \circ \frac{1}{N_v}\sum_n\phi_{v,n}$.

% GRTM is an end-to-end model, using the evidence of observed links to supervise the model in such a way that the contents of the shared images are statistically connected to the links between users. Such evidence can then guide the model towards better predictions. 
% It should be noted that although during the training process evidence of observed links are used for supervision, the method is very different from graph-based method such as collaborate filtering. During prediction, no social graph information is needed, and hence our method is also free from sparsity and cold start problem commonly shown in graph-based methods.

\section{Experimental Results}
% In this section, the dataset, the experiments and the results are discussed.
% Our work focuses on discovering connections among users by using the images they have shared on social media.
% The experiment implements the proposed GRTM and trains the model using an image dataset from a social network site. 
In this section, experiments are conducted to investigate the effectiveness of the proposed GRTM for connection discovery using user shared images.
%======================================================
\subsection{Dataset and Experimental Setup}
%=======================================

% put the experiment procedure as a figure.
% The setting of the experiment is shown in Fig. \ref{fig:settings}.
A set of 201,006 images uploaded by 542 users are scraped from Flickr, an online social network for image sharing, with millions of images uploaded. The 542 users are selected randomly from images under the same tag query page to provide diversity.
The average number of shared images for each user is 370, covering diverse content, and there are 902 connections among the 542 users.
% The average number of friends of a user is 170, for which there are 902 connections among the 542 users.
% The goal of the experiment is to infer these connections using the set of images uploaded by the users, even without access to the social graph.
The images are processed by GoogLeNet \cite{GoogLeNet2015}, pretrained using the ILSVRC 2014 dataset, for image representation, so that rich semantic information can be extracted. 
% For the training process of the proposed GRTM, all the user shared images are observed. 
The links between users are divided into two parts: 60\% of the links are used for training and 40\% of the links are used for testing. Using the user shared images and training links, the latent variables and model parameters of the GRTM are estimated through an iterative process, as described in the previous section. 
% Only the training positive links are considered observed during training process and all rest of the links (testing positive links and all negative links) are considered as unobserved. 
Using Eq. \ref{eq:predict}, the testing process is then undertaken to predict the probability of the links existing between users , excluding the observed training links, and the results are then compared with the groundtruth of the testing links. The Dirichlet hyperparameter $\alpha$ is set to 2.0,
% Since the topics covered in one user's collection of images are usually diverse, the Dirichlet hyperparameter $\alpha$ is set to be larger than 1.0, at 2.0 in our experiment. However, it is observed that the performance is similar with $\alpha$ ranges from 2.0 to 5.0. 
% Generally for random initialization it is more appropriate to pose an additional prior on Gaussian topic $\{\mu,\Sigma\}_{1:K}$. Instead, for faster convergence, we initialize the topic parameters using the clustering result of $k$-means, which is common for Gaussian mixture model.
and the number of topics is set to be 100 in our experiment. Though the number of topics has some influence to some extent, the overall performance does not vary much.
% \begin{figure*}
%   \centering
%   \includegraphics[width=6.0in]{image/auc_v4.pdf}
%   \caption{Performance comparison of Mean method \cite{zhuang2011modeling}, BoFT method \cite{cheung2015connection} and proposed GRTM: (a) ROC curve, (b) Precision-Recall curve.}
%   \label{fig:resultPrecision}
% \end{figure*}

As a comparison with previous works on similar applications, the Mean method from \cite{zhuang2011modeling} where users' profiles are constructed through their shared images by taking the mean features of all the images, and the BoFT method \cite{cheung2015connection} where users' profiles are obtained by counting the occurrences of cluster labels in users' collections obtained through image clustering, are implemented. 
% For Mean method from \cite{zhuang2011modeling}, the user profile is constructed through their shared image by taking the mean feature of all the image features. For Bag-of-Feature Tagging (BoFT) method from \cite{cheung2015connection}, the user profile is obtained by counting the occurrences of cluster labels in users' collections obtained through image clustering. 
The prediction of user connections is thus conducted through computing the similarity between users. 
% As with the GRTM experiment, training links are excluded during evaluation 
And instead of using the scale invariant feature transform (SIFT) for image feature extraction, CNN (GoogLeNet) features are used.
% , since it is shown in a number of prior works \cite{lecun2015deep}\cite{Decaf2013} that features extracted from CNN are more effective than conventional features. 

%======================================
%Talk about the setting of each technique
%=============================
% \begin{figure}[t]
%   \centering
%   %\vspace{-10pt}
%   \includegraphics[width=3.5in]{image/iteration.pdf}
%   \caption{Training iteration process.}
%   \label{fig:iteration}
% \end{figure}

%===============================

\subsection{Results}
\begin{figure}[t]
  \centering
  \includegraphics[width=3.0in]{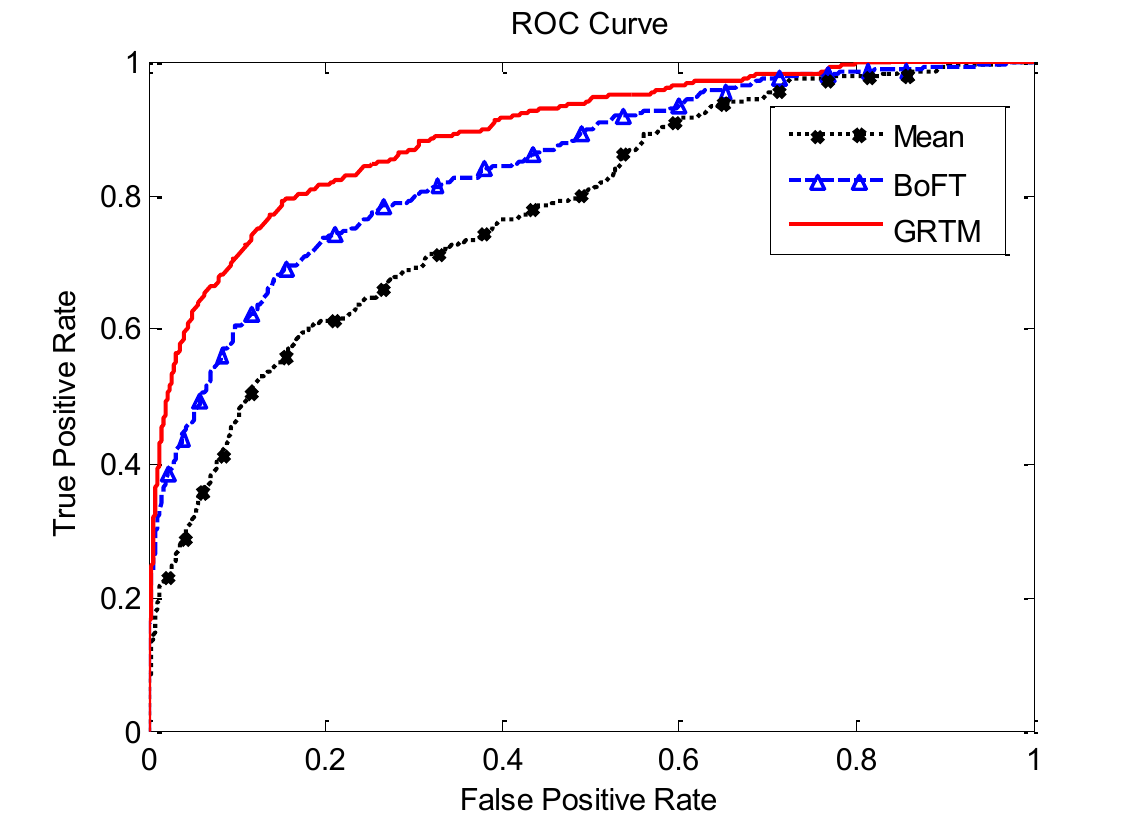}
  \caption{ROC curve for Mean method \cite{zhuang2011modeling}, BoFT method \cite{cheung2015connection} and proposed GRTM.}
  \label{fig:resultroc}
\end{figure}
\begin{figure}[t]
  \centering
  \includegraphics[width=3.0in]{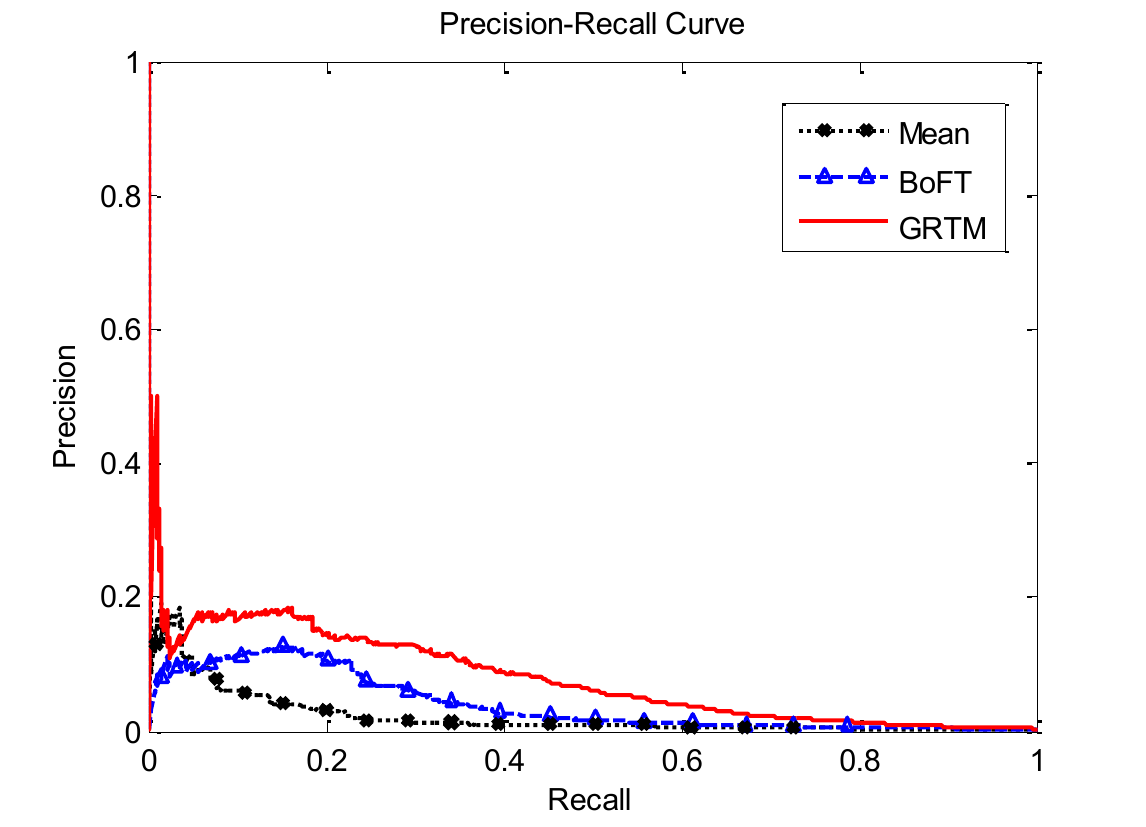}
  \caption{Precision-Recall curve for Mean method \cite{zhuang2011modeling}, BoFT method \cite{cheung2015connection} and proposed GRTM.}
  \label{fig:resultpr}
\end{figure}

%\subsubsection{ROC and Precision-recall Curve}

% The training iteration process of the proposed GRTM with above setting is shown in Fig. \ref{fig:iteration}, which plots the evidence lower bound of the log likelihood as in Eq. \ref{eq:elbo} with respect to the number of iterations. As it is shown, with a good intialization by the result of $k$-means the evidence lower bound quickly converges to the maximum. After around 30 iterations, the objective is already very close to the maximum.

The prediction performance of the Mean method, BoFT method and proposed GRTM is shown in Fig. \ref{fig:resultroc}, where both the receiver operating characteristic (ROC) curve and precision-recall curve are plotted. The ROC curve is the function of true positive rate (TPR) and false positive rate (FPR) and is used to measure how well the model can distinguish whether two users are friends or not. As it is shown, the ROC curve of the GRTM clearly dominates the other two methods. For example, for the point in the ROC curve with TPR of 0.8, given a pair of users with a true friendship, GRTM has 80\% probability of predicting it correctly, while given a pair of users without a friendship, it has 20\% probability of predicting it incorrectly. For the same TPR, however, the BoFT method has 35\% probability of incorrect prediction for users without a friendship, and Mean method has 45\%. Overall, the area under curve (AUC) for ROC curve of the proposed GRTM is 0.89, while that of the BoFT method is 0.84 and that of Mean method is 0.78.
As an alternative measure, the precision-recall curve, as shown in Fig. \ref{fig:resultpr}, is the function of the precision rate and recall rate. It measures what fraction of the recommended candidates are the user's true friends and what fraction of true friends are recommended. As with the ROC curve, the precision-recall curve of the proposed GRTM also dominates the others. For example, for the same recall rate of 10\%, the precision rate of the GRTM is 18\%, while that of the BoFT method is 11\% and that of the Mean method is 5.9\%. Overall, the precision-recall AUC of the proposed GRTM is 0.05, while that of the BoFT method is 0.039 and that of the Mean method is 0.021.

\begin{figure}[t]
  \centering
  %\vspace{-10pt}
  \includegraphics[width=3.5in]{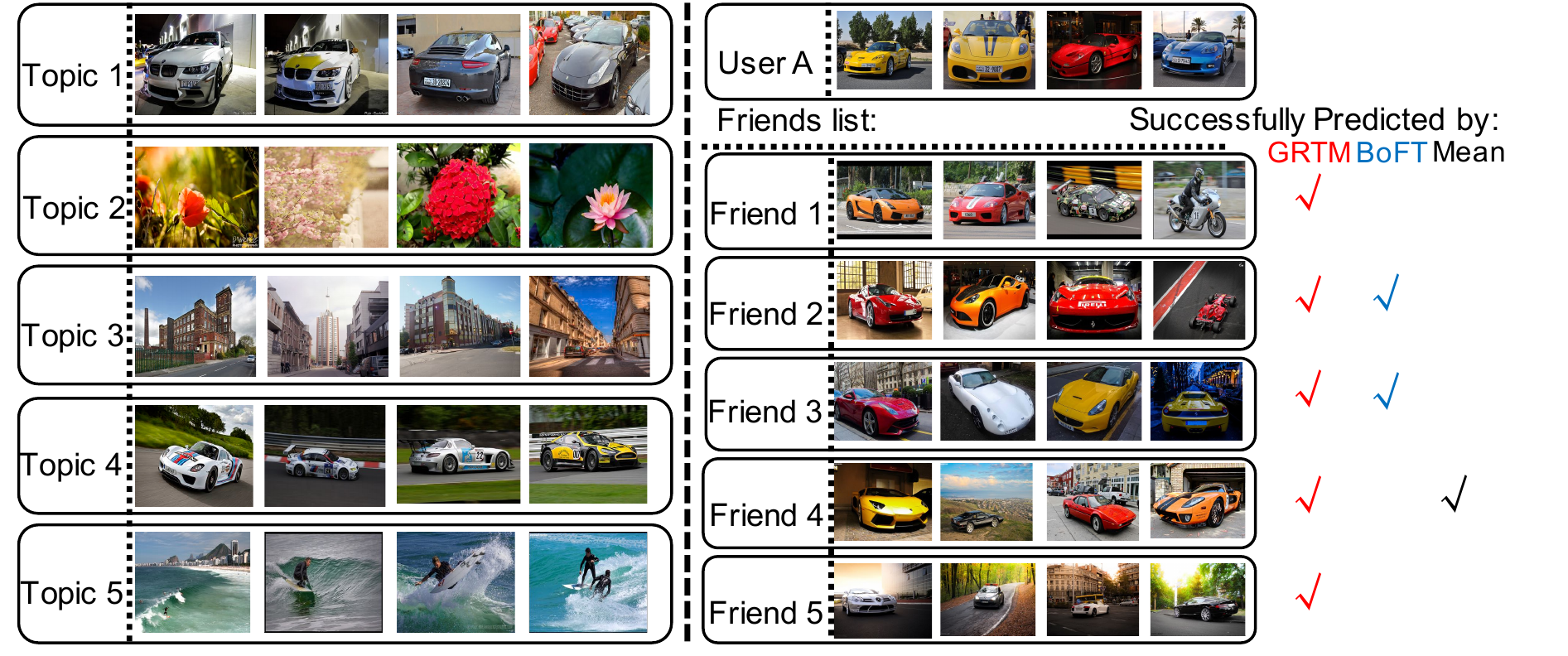}
  \caption{Left: examples of Gaussian topics; Right: examples of friendship prediction by Mean method, BoFT method and the proposed GRTM.}
  \label{fig:gaussianExample}
\end{figure}

In order to illustrate how the proposed GRTM interprets the user shared images, some example images are shown in Fig. \ref{fig:gaussianExample}. Since each Gaussian topic is defined by $\mu_k$ and $\Sigma_k$ and each image has a probability under the Gaussian topic, images that achieve the highest probabilities under that topic are selected as representatives. As shown in Fig. \ref{fig:gaussianExample}, the semantic meaning of the Gaussian topics, such as cars, flowers, buildings, race cars and surfing, can be readily distinguished. The Gaussian topics cover a variety of objects, scenes, sports, etc. Those images that contain more than one topic could also be revealed through the probabilities under different Gaussian topics. It is also noted that ordinary cars and race cars are separated as different topics, which is also beneficial to reflect users' preferences for subsequent connection discovery.
Fig. \ref{fig:gaussianExample} shows examples of the friendship prediction results obtained by the three methods. The example user, $A$, shares a lot of car images. For recommendation purposes, all three methods are directed to recommend 10 friends. As shown in the figure, out of the 10 recommendations, the proposed GRTM successfully predicts 5 of them, whereas the BoFT only successfully predicts 2 and the Mean method successfully predicts 1. Furthmore, the successful predictions of the other two methods are in fact a subset of those of the proposed GRTM. 
% Due to space limitation, other recommendations that are incorrectly predicted by all three methods are not shown.

% In summary, the proposed GRTM, which closely relates user shared images to user connections explicitly, achieves significant better performance compared with previous connection discovery works that seperate the modeling of users' interests from the links between users. 
%The proposed end-to-end model simutaneously considers the generative process of image content and links between users in a coherent way and provides a systematic method to close the gap between the two.
%The results will be discussed in the discussion section.
% \begin{figure}[t]
%   \centering
%   %\vspace{-10pt}
%   \includegraphics[width=2.5in]{image/example_gaussian_topic.pdf}
%   \caption{Examples of Gaussian topics.}
%   \label{fig:gaussianExample}
% \end{figure}
% \begin{figure}[t]
%   \centering
%   %\vspace{-10pt}
%   \includegraphics[width=2.5in]{image/prediction_example_v4.pdf}
%   \caption{Examples of friendship prediction by Mean method, BoFT method and the proposed GRTM.}
%   \label{fig:predictExample}
% \end{figure}

\section{Conclusion}
This paper proposes a Gaussian relational topic model (GRTM) for connection discovery using user shared images in social media. The GRTM not only models users' interests as latent variables through user shared image content but also models the connections between users as a result of their shared images. It explicitly relates user shared images to the connections between users in a hierarchical, systematic and supervisory way and provides an end-to-end model for connection discovery using shared images.
It is demonstrated by experiment that the proposed model significantly outperforms the methods in previous works where the modeling of users' interests and connection discovery are separated.


\section*{Acknowledgment}
This work is supported by the HKUST-NIE Social Media Lab., HKUST.



%\bibliographystyle{IEEEtran}


% \end{thebibliography}

\bibliographystyle{IEEEtran}
\bibliography{IEEEabrv,sigproc}

\end{document}